\documentclass{emulateapj}
\usepackage{apjfonts}
\newcommand{\kel}{\mbox{ K}}
\newcommand{\mkel}{\mbox{ mK}}

\newcommand{\Mpc}{\mbox{ Mpc}}

\newcommand{\msun}{\mbox{ M$_\odot$}}

\newcommand{\hunits}{\mbox{ km s$^{-1}$ Mpc$^{-1}$}}
\newcommand{\bq}{\begin{equation}}
\newcommand{\eq}{\end{equation}}
\newcommand{\bqa}{\begin{eqnarray}}
\newcommand{\eqa}{\end{eqnarray}}

\newcommand{\bxh}{\bar{x}_H}
\newcommand{\xh}{x_H}
\newcommand{\bxi}{\bar{x}_i}
\newcommand{\Qb}{\bar{Q}}
\newcommand{\xixx}{\xi_{xx}}
\newcommand{\xidd}{\xi_{\delta \delta}}
\newcommand{\xixd}{\xi_{x\delta}}
\newcommand{\Pxx}{P_{xx}}
\newcommand{\Pdd}{P_{\delta \delta}}
\newcommand{\Pxd}{P_{x\delta}}
\newcommand{\br}{{\bf r}}
\newcommand{\mpix}{m_{\rm pix}}

\begin{document}

\title{The Growth of \ion{H}{2} Regions During Reionization}

\author{Steven R.  Furlanetto\altaffilmark{1}, Matias
Zaldarriaga\altaffilmark{2,3}, \& Lars Hernquist\altaffilmark{2}}

\altaffiltext{1} {Division of Physics, Mathematics, \& Astronomy;
  California Institute of Technology; Mail Code 130-33; Pasadena, CA
  91125; sfurlane@tapir.caltech.edu}

\altaffiltext{2} {Harvard-Smithsonian Center for Astrophysics, 60
Garden St., Cambridge, MA 02138}

\altaffiltext{3} {Jefferson Laboratory of Physics, Harvard University, 
Cambridge, MA 02138}

\begin{abstract}

Recently, there has been a great deal of interest in understanding the
reionization of hydrogen in the intergalactic medium (IGM).  One of
the major outstanding questions is how this event proceeds on large scales.
Motivated by numerical simulations, we develop a model for the growth
of \ion{H}{2} regions during the reionization era.  We associate
ionized regions with large-scale density fluctuations and use the
excursion set formalism to model the resulting size distribution.  We
then consider ways in which to characterize the morphology of ionized
regions.  We show how to construct the power spectrum of fluctuations
in the neutral hydrogen field.  The power spectrum contains definite
features from the \ion{H}{2} regions which should be observable with
the next generation of low-frequency radio telescopes through surveys
of redshifted 21 cm emission from the reionization era.  Finally, we
also consider statistical descriptions beyond the power spectrum and
show that our model of reionization qualitatively changes the
distribution of neutral gas in the IGM.

\end{abstract}
  

\section{Introduction}
\label{intro}

The reionization of the intergalactic medium (IGM) is one of the
landmark events in early structure formation.  It marks the epoch at
which radiative feedback from luminous objects impacted the farthest
reaches of the Universe -- the point at which structure formation
affected every baryon in the Universe, albeit indirectly.  The timing,
duration, and character of this event contain an enormous amount of
information about the first cosmic structures and also have important
implications for later generations of baryonic objects.  For these
reasons, a great deal of attention -- both observational and
theoretical -- has recently been focused on this process.  Most
significantly, reionization is an important signpost that connects
several disparate observations.  Currently the data provide tantalizing
hints about the ionization history of the Universe but few definitive
answers.  Observations of Ly$\alpha$ absorption in the spectra of
high-redshift quasars indicate that reionization ends at $z \sim 6.5$
\citep{becker,fan,white03,wyithe04-prox}, although this interpretation
is controversial \citep{songaila04}.  The main difficulty with these
measurements is that the Ly$\alpha$ optical depth is extremely large
in a fully neutral medium \citep{gp}, making it difficult to place
strong constraints when the neutral fraction exceeds $\sim 10^{-3}$.
On the other hand, measurements of the cosmic microwave background
(CMB) imply a high optical depth to electron scattering, apparently
requiring reionization to begin at $z \ga 14$
\citep{kogut03,spergel03}.  Unfortunately, the CMB data provide only
an integral constraint on the ionization history.  Taken together,
these observations rule out simple pictures of fast reionization
(e.g., \citealt{barkana01} and references therein), but it is not yet
clear what they do imply about early generations of luminous sources
\citep{sokasian03a,sokasian03b,wyithe03,cen03,haiman03,onken03,fukugita03}.

At the same time, our theoretical understanding of how reionization
proceeds, given some source population, has been advancing rapidly.
Most models of reionization are based on the growth of \ion{H}{2}
regions around individual galaxies \citep{arons72,barkana01}.  These
models use semi-analytic techniques to compute the evolution of global
quantities like the total ionized fraction.  However, they are unable
to describe the morphology of reionization, an issue that is both
theoretically interesting and observationally accessible.  Morphology
is a difficult aspect to address because it depends on such factors as
the locations of individual sources, the clumpiness of the IGM, the
underlying density distribution (the cosmic web), recombinations, and
radiative transfer effects.  Some semi-analytic models have been
developed to address these issues.  The most popular assumes that
reionization is controlled by recombinations and proceeds from low to
high density regions \citep{miralda00}, but these models are
approximate at best.  Fortunately, it has recently become possible to
incorporate radiative transfer into numerical simulations of the
reionization era
\citep{gnedin00,sokasian03a,sokasian03b,ciardi03-sim}, at least as a
post-processing step.  Because simulations can include all of the
above processes (with the possible exception of clumpiness, which
still requires extremely high mass resolution), they give a more
nuanced view of the reionization process.  One of the chief lessons of
the simulations is that reionization is significantly more
inhomogeneous than expected.  The classical picture of a large number
of small \ion{H}{2} regions surrounding individual galaxies does not
match the simulations well; instead, a relatively small number of
large ionized regions appear around clusters of sources (see, for
example, Figure 6 of \citealt{sokasian03a}).  Moreover, in the
simulations reionization proceeds from high to low density regions,
implying that recombinations play only a secondary role in
determining the morphology of reionization; instead, large-scale bias
plays a dominant role.  This picture suggests a new approach to
analytic models of reionization, one that takes into account the
large-scale density fluctuations that are ultimately responsible for
this ionization pattern.  We describe such a model in \S
\ref{bubbles}.  We derive the size distribution of \ion{H}{2} regions
in a way analogous to the \citet{press} halo mass function and show
that it has the qualitative features seen in the simulations.

Observing the morphology of \ion{H}{2} regions requires new
observational techniques as well as robust ways to characterize the
data.  Among the most exciting approaches to study the reionization
process are surveys of 21 cm emission from neutral hydrogen at high
redshifts \citep{field58,scott,mmr,zald04}.  The idea behind these
observations is to map the fluctuating pattern of emission (or
absorption) from neutral hydrogen in the Universe over a range of
frequencies.  This yields a measurement of fluctuations due to both
the density field and the \ion{H}{2} regions.  Because 21 cm emission
comes from a single spectral line, such observations allow a
three-dimensional reconstruction of the evolution of neutral hydrogen
in the IGM and, owing to the design of low-frequency radio
telescopes, can probe the large cosmological volumes needed to study
the IGM.  Despite numerous technical challenges, instruments able to
make the necessary observations will be built in the coming years.
These include the Primeval Structure Telescope
(\emph{PAST}),\footnote{ See
http://astrophysics.phys.cmu.edu/$\sim$jbp for details on PAST.} the
Low Frequency Array (\emph{LOFAR}),\footnote{ See http://www.lofar.org
for details on LOFAR.} and the Square Kilometer Array
(\emph{SKA}).\footnote{ See http://www.skatelescope.org for details on
the SKA.}  The major obstacles to high-redshift 21 cm observations are
the many bright foreground sources, which include Galactic synchrotron
emission, free-free emission from galaxies \citep{oh03}, faint
radio-loud quasars \citep{dimatteo02}, and synchrotron emission from
low-redshift galaxy clusters \citep{dimatteo04}.  Fortunately, all of
these foregrounds have smooth continuum spectra.  Zaldarriaga et
al. (2004; hereafter ZFH04) showed that the foregrounds can be removed
to high precision because the 21 cm signal itself is uncorrelated over
relatively small frequency ranges.  ZFH04 also showed how to compute
the angular power spectrum of the 21 cm sky as a function of
frequency, given some model of reionization.  This is the simplest
statistical measure of the morphology of \ion{H}{2} regions, and they
showed that it is quite powerful in distinguishing different stages of
reionization (at least in a simple toy model).
  
Of course, predictions for the 21 cm signal, or any other measurement
of reionization, rely on an accurate model for reionization.  The
initial conditions are straightforward: when the neutral fraction
$\bxh \approx 1$, the power spectrum follows that of the density
\citep{tozzi}.  This breaks down when the \ion{H}{2} regions appear,
and more sophisticated approaches become necessary.  There have been
several recent attempts to predict the signal using numerical
simulations \citep{ciardi03,furl-21cmsim,gnedin03}.  However, the
boxes in these simulations have sizes $\la 20 \Mpc$, subtending only a
few arcminutes on the sky.  Because the angular resolution of low
frequency radio telescopes at the required sensitivity is also a few
arcminutes (ZFH04), such predictions require extrapolations to larger
scales, which may be dangerous given the already large sizes of
ionized regions.  With our analytic model for the size distribution of
these bubbles, we are able to make the first detailed statistical
characterizations of the fluctuating ionization pattern on the large
scales most relevant to observations.  We begin with the power
spectrum as the simplest description.  In \S \ref{ps}, we show how to
compute the power spectrum for an arbitrary size distribution, and in
\S \ref{res} we apply that formalism to our model of \ion{H}{2}
regions.

The power spectrum is only one way to describe the 21 cm field.  It is
an excellent approximation if fluctuations in the matter density
dominate the signal, because the density field is neary gaussian on
the large scales of interest here.  Unfortunately, once the \ion{H}{2}
regions dominate the power spectrum, the probability distribution is
no longer gaussian \citep{morales03}.  There have, however, been no
attempts to describe the distribution.  Using our model for
reionization, we discuss some of the relevant features in \S
\ref{nongauss}.  We show that the reionization model qualitatively
changes the character of fluctuations in the neutral gas density.

This paper is primarily concerned with developing a general and useful
model for the morphology of reionization and with the crucial features
of that model.  We will focus here on the physics of reionization
rather than on their observable consequences; we consider some of
these in a companion paper about the 21 cm signal expected from high
redshifts \citep{furl04b}.

In our numerical calculations, we assume a cosmology with
$\Omega_m=0.3$, $\Omega_\Lambda=0.7$, $\Omega_b=0.046$, $H=100 h
\hunits$ (with $h=0.7$), $n=1$, and $\sigma_8=0.9$, consistent with
the most recent measurements \citep{spergel03}.

\section{\ion{H}{2} Regions During Reionization}
\label{bubbles}

In this section, we will calculate the size distribution of ionized
regions within the IGM.  So long as ultraviolet photons are
responsible for reionization, it is a good approximation to divide the
IGM into a fully neutral component inside of which sit discrete, fully
ionized bubbles.  We will make this approximation throughout; we
consider some more complicated cases in \citet{furl04b}.  A complete
description requires accurately locating the ionizing sources, fully
resolving the clumpiness of the IGM, following recombinations, and
performing radiative transfer.  These problems all require numerical
simulations (e.g., \citealt{gnedin00,sokasian03a,sokasian03b,
ciardi03-sim} and references
therein).  As a result, semi-analytic models have avoided this
question and focused on global quantities such as the mean neutral
fraction $\bxh$.  The simplest semi-analytic model, and the usual
assumption in the literature, is to associate each ionized region with
a single galaxy (e.g., \citealt{barkana02-bub,loeb04}).  In this case
the size distribution $dn/dm$ (where $m$ is the mass of the ionized
region) follows directly from the halo mass function if we make the
simple ansatz that
\bq
m_{\rm ion} = \zeta m_{\rm gal},
\label{eq:zeta}
\eq
where $m_{\rm gal}$ is the mass in a collapsed object and $\zeta$ is
some efficiency factor.  It could, for example, be written $\zeta =
f_{\rm esc} f_\star N_{\gamma/b} n_{\rm rec}^{-1}$, with $f_{\rm esc}$
the escape fraction of ionizing photons from the object, $f_\star$ the
star formation efficiency, $N_{\gamma/b}$ the number of ionizing
photons produced per baryon in stars, and $n_{\rm rec}$ the typical
number of times a hydrogen atom has recombined.  These parameters all
depend on the uncertain source properties and can be functions of
time; we will consider several possible values for $\zeta$ below.

Because the mass function is steep at high redshifts, the resulting
\ion{H}{2} regions are quite small (see the discussion of Figure
\ref{fig:dndr-comp} below).  This conflicts with even the most basic
pictures from simulations \citep{sokasian03a,ciardi03-sim}.
``Typical'' ionized regions in simulations extend to several comoving
Mpc in radius even early in overlap, many times larger than Str{\"
o}mgren spheres around individual galaxies.  The reason is simply that
the Str{\" o}mgren spheres of nearby protogalaxies add, so that biased
regions tend to host surprisingly large ionized regions.  For example,
Figure 6 of \citet{sokasian03a} shows that \ion{H}{2} regions tend to
grow around the largest clusters of sources, in this case primarily
along filaments.  In fact, the radius of the \ion{H}{2} regions
quickly exceeds the correlation length of galaxies, so it is difficult
to see how to construct a model for the bubbles based on ``local''
galaxy properties.

Therefore, in order to describe the neutral fraction
field, $\xh$, we need to take into
account large-scale fluctuations in the density field.  Here we
describe a simple way to do so.  We again begin with the ansatz of
equation (\ref{eq:zeta}) and ask whether an isolated region of mass
$m$ is \emph{fully} ionized or not.  Because it is isolated, the
region must contain enough mass in luminous sources to ionize all of
its hydrogen atoms; thus we can impose a condition on the collapse
fraction:
\bq
f_{\rm coll} \ge f_x \equiv \zeta^{-1}.  
\label{eq:fcollcond}
\eq
In the extended Press-Schechter model \citep{bond91,lacey}, the collapse
fraction is a deterministic function of the mean linear overdensity
$\delta_m$ of our region:
\bq
f_{\rm coll} = {\rm erfc} \left[ \frac{\delta_c(z) -
    \delta_m}{\sqrt{2[\sigma^2_{\rm min} - \sigma^2(m)]}} \right],
\label{eq:fcoll}
\eq 
where $\sigma^2(m)$ is the variance of density fluctuations on the
scale $m$, $\sigma^2_{\rm min}=\sigma^2(m_{\rm min})$, $\delta_c(z)$
is the critical density for collapse, and $m_{\rm min}$ is the minimum
mass of an ionizing source\footnote{Note that in equation
  (\ref{eq:fcoll}) the growth of structure is encoded in the time
  evolution of $\delta_c(z)$, with $\sigma^2(m)$ constant in time. 
We adopt this convention in the rest of the paper.}.  
Unless otherwise specified, we will take $m_{\rm min}$ to
be the mass corresponding to a virial temperature of $10^4 \kel$, at
which atomic hydrogen line cooling becomes efficient.  Note that this
expression assumes that the mass fluctuations are gaussian on the
scale $m$; the formula thus begins to break down when we consider mass
scales close to the typical size of collapsed objects.  Armed with
this result, we can rewrite condition (\ref{eq:fcollcond}) as a
constraint on the density:
\bq \delta_m \ge \delta_x(m,z) \equiv
\delta_c(z) - \sqrt{2} K(\zeta) [\sigma^2_{\rm min} -
\sigma^2(m)]^{1/2},
\label{eq:deltax}
\eq
where $K(\zeta) = {\rm erf}^{-1}(1 - \zeta^{-1})$.  We see that
regions with sufficiently large overdensities will be able to
``self-ionize.''  

In order to compute the size distribution of ionized regions we must
overcome two additional, but related, difficulties.  First, we
apparently must settle on an appropriate smoothing scale $m$.  Second,
we must take into account ionizing photons from galaxies
\emph{outside} of the region under consideration.  In other words, an
underdense void $m_1$ may be ionized by a neighboring cluster of
sources in an overdense region $m_2$ provided that the cluster has
enough ``extra'' ionizing photons.  But notice that we can solve the
latter problem by changing our smoothing scale to $m_1+m_2$: then the
net collapse fraction in this region would be large enough to
``self-ionize.''

This suggests that we wish to assign a point in space to an ionized
region of mass $m$ if and only if the scale $m$ is the \emph{largest}
scale for which condition (\ref{eq:deltax}) is fulfilled.  If this
procedure can be done self-consistently, we will not need to
arbitrarily choose a smoothing scale.  Our problem is analogous to
constructing the halo mass function through the excursion set
formalism \citep{bond91}: starting at $m=\infty$, we move to smaller
scales surrounding the point of interest and compute the smoothed
density field as we go along.  Once $\delta_m=\delta_x(m,z)$, we have
identified a region with enough sources to ionize itself, and we
assign these points to objects of the appropriate mass.  To obtain the
mass function, we need to find the distribution of first up-crossings
above the curve described by $\delta_x$.  (We are concerned only with
the first-crossing distribution because those trajectories that later
wander below the barrier correspond to regions ionized by sources in
neighboring volumes.)  Again, we need not choose a smoothing scale;
each point is \emph{assigned} to an object of mass $m$ based on its
own behavior.

\begin{figure}
\plotone{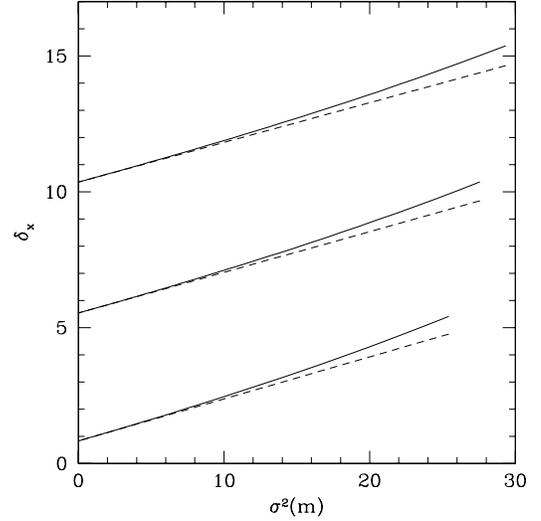}
\caption{The density threshold $\delta_x(\sigma^2,z)$ at several
  different redshifts, assuming $\zeta=40$. The curves are for
  $z=20,16$, and $12$, from top to bottom.  Within each set, the solid
  curve is the true $\delta_x(m,z)$ and the dashed line is the fit
  $B(m,z)$.}
\label{fig:dx}
\end{figure}

The solid lines in Figure \ref{fig:dx} show the barrier
$\delta_x(m,z)$ for several redshifts as a function of $\sigma^2(m)$.
In each case the curves end at $\sigma^2(\zeta m_{\rm min})$; this is
the minimum size of an \ion{H}{2} region in our formalism.  The Figure
shows an important difference between our problem and the excursion
set formalism applied to the halo mass function.  In the latter case,
the barrier $\delta_c(z)$ is independent of mass.  Clearly this would
not be a good approximation in our case.  Unfortunately, there is no
general method for constructing the first-crossing distribution above
a barrier of arbitrary shape (but see Sheth \& Tormen [2002] for an
approximate method).  The most complicated case for which an analytic
solution is available is a linear barrier \citep{sheth98}.  The dashed
curves in Figure \ref{fig:dx} show linear ``fits'' to the barrier
constructed in the following way.  First note that, as $m \rightarrow
\infty$,
\bq
\delta_x \rightarrow B_0 \equiv \delta_c(z) - \sqrt{2} K(\zeta)
\sigma_{\rm min}.
\label{eq:b0}
\eq
Also, at any given $\sigma^2$, the slope is simply
\bq
\frac{\partial \delta_x}{\partial \sigma^2} = \frac{
  K(\zeta)}{\sqrt{2(\sigma^2_{\rm min} - \sigma^2)}}.
\label{eq:b1}
\eq
We define $B_1$ to be this slope evaluated at $\sigma^2=0$.  The
dashed lines in Figure \ref{fig:dx} are $B(m,z) = B_0 + B_1
\sigma^2(m)$, i.e. a linear fit to the true barrier at $m = \infty$.
We see that this is a reasonable approximation to the true barrier
shape for $\sigma^2$ that are not too large.  The fit departs from
$\delta_x$ as the mass approaches the size of \ion{H}{2} regions
around individual galaxies.  However, equation (\ref{eq:fcoll}) --
upon which the entire approach is predicated -- also breaks down on
small mass scales.  Thus we do not consider it necessary to improve
the fit.  (In any case, changing the slope of the barrier to fit
$\sigma^2(\zeta m_{\rm min})$ exactly does not significantly change
our results except at early times, when the bubbles are still quite
rare.)

\begin{figure}
\plotone{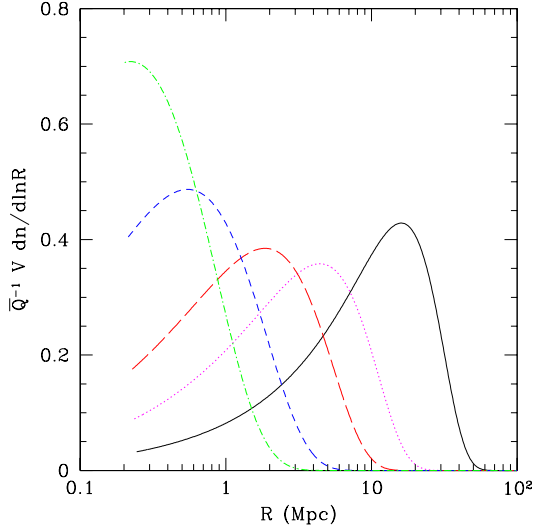}
\caption{ The bubble size distribution $\Qb^{-1} V dn/d \ln R$ at
  several different redshifts in our model, assuming $\zeta=40$ (note
  that $R$ is the comoving size).  Dot-dashed, short-dashed,
  long-dashed, dotted, and solid lines are for $z=18,\,16,\,14,\,13$,
  and $12$, respectively. These have $\bar{Q} = 0.037,0.11,0.3,0.5$,
  and $0.74$.}
\label{fig:dndr}
\end{figure}

\begin{figure}
\plotone{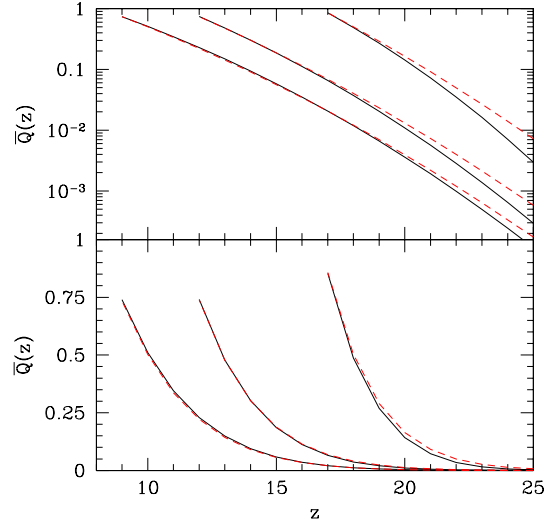}
\caption{The global ionization history for several scenarios.  The
  curves that rise from zero assume $\zeta=500,\,40$, and $12$, from
  right to left; solid lines are for our model and dashed lines are the
  ``true'' values, $\zeta f_{\rm coll}$.}
\label{fig:qbar}
\end{figure}

The advantage of a linear fit is that we can now write the mass
function analytically \citep{sheth98}: 
\bq
m \frac{dn}{dm} = \sqrt{\frac{2}{\pi}} \ \frac{\bar{\rho}}{m} \ \left|
  \frac{d \ln \sigma}{d \ln m} \right| \ \frac{B_0}{\sigma(m)} \exp
  \left[ - \frac{B^2(m,z)}{2 \sigma^2(m)} \right].
\label{eq:dndm}
\eq
This is the comoving number density of \ion{H}{2} regions with masses
in the range $(m,m+dm)$.  Figure \ref{fig:dndr} shows the resulting
size distributions at several redshifts for $\zeta=40$.  The
dot-dashed, short-dashed, long-dashed, dotted, and solid curves
correspond to $z=18,\,16,\,14,\,13$, and $12$, respectively.  The
curves begin at the radius corresponding to an \ion{H}{2} region
around a galaxy of mass $m_{\rm min}$.  We have normalized each curve by
the fraction of space $\Qb$ filled by the bubbles,
\bq
\Qb = \int dm \, \frac{dn}{dm} \, V(m),
\label{eq:qbar}
\eq
where $V(m)$ is the comoving volume of a bubble of mass $m$.  We show
the evolution of $\Qb$ for several choices of $\zeta$ by the solid
lines in Figure \ref{fig:qbar}; the curves in Figure \ref{fig:dndr}
range from $\Qb=0.037$ to $\Qb=0.74$.  When the ionized fraction is
small, the ionized regions are also small, with characteristic sizes
$\la 0.5 \Mpc$.  At this point they are not much bigger than the
Str{\" o}mgren spheres surrounding individual galaxies.  However, the
size increases rapidly as the neutral fraction decreases; when
$\Qb=0.5$, the bubbles are already several Mpc in size.  The
characteristic scale then begins to increase extremely rapidly because
$B_0 \rightarrow 0$ as we approach overlap (see Figure \ref{fig:dx}).
This behavior matches the results of the simulations cited above --
although note that the scales we find can exceed the simulation box
sizes well before overlap (see below).

This contrasts sharply with a scenario in which we assign ionized
regions to individual galaxies: the top panel of Figure
\ref{fig:dndr-comp} compares the predictions of our model with one in
which each galaxy hosts its own distinct ionized bubble.  Note that we
have not normalized the curves by $\Qb$.  In the galaxy-based model,
we see that the bubble sizes change only very slowly; the filling
factor is dominated by the smallest galaxies.  In such a scenario,
overlap is achieved not by the growth of existing \ion{H}{2} regions
but through the formation of more distinct bubbles.  Also note that,
provided our model is correct, the sizes of ionized regions cannot be
determined even from the Str{\" o}mgren spheres around ``large'' or
$L_*$ galaxies.  Instead large-scale fluctuations are \emph{required}
in order to understand the bubble pattern.  The disparity between the
models clearly increases rapidly as we approach overlap, but it is
significant even at early times.

\begin{figure}
\plotone{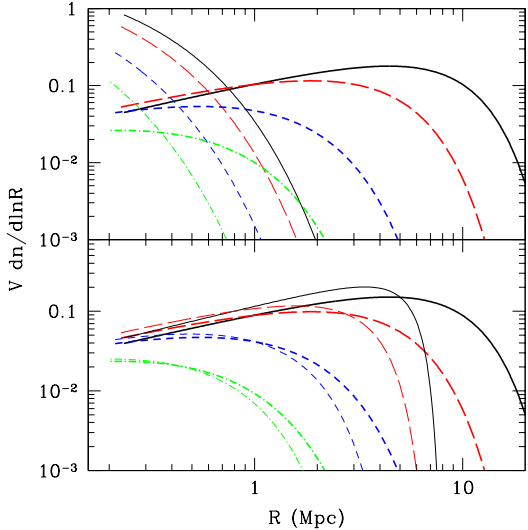}
\caption{ The bubble size distribution $V dn/d \ln R$ at several
  different redshifts, assuming $\zeta=40$.  In both panels, the thick
  lines show our model with dot-dashed, short-dashed, long-dashed, and
  solid lines corresponding to $z=18,\,16,\,14$, and $13$,
  respectively.  In the top panel, the thin lines assume individual
  galaxies source distinct \ion{H}{2} regions.  In the bottom panel,
  the thin lines use equation (\ref{eq:dndmbox}), with $L_{\rm
  box}=10h^{-1} \Mpc = 14.3 \Mpc$.  }
\label{fig:dndr-comp}
\end{figure}

The mass function of ionized regions has qualitatively different
behavior from the usual halo mass function.  The most significant
difference is that it has both low and high mass cutoffs.  This occurs
because the barrier rises more steeply than $\sigma(m)$, and it
becomes harder for a trajectory to cross it at small masses
\citep{sheth02}.\footnote{Note that the real barrier rises more
quickly than our linear approximation. This will make the low mass
cutoff occur slightly earlier, increasing the sharpness of the peak in
the size distribution.}  As a result, the mass function has a
characteristic scale that becomes sharper as $\Qb$ increases.  In
general, our model is good news for attempts to constrain the
\ion{H}{2} regions observationally.  It predicts large features with a
characteristic scale, which should make the ionized bubbles easier to
identify.

Although we have argued that the general properties of our model
reproduce the qualitative features found in simulations, we have not
performed a quantitative comparison.  This will clearly require a
large simulation box in order to reproduce the size distribution
we expect.  In our model, the bubble sizes are essentially set by
large-scale fluctuations of the density field; as emphasized by
\citet{barkana03}, the finite simulation size will bias these
fluctuations and reduce the characteristic sizes.  A numerical
simulation typically forces $\delta=0$ on the size scale of the box,
$L_{\rm box}$.  We can calculate the expected size distribution in a
box of the appropriate size by changing the origin of the excursion
set formalism from $(\sigma^2=0,\delta=0)$ to $(\sigma^2=\sigma^2_{\rm
box},\delta=0)$.  The results are exactly analogous to the extended
Press-Schechter formalism for calculating conditional halo mass
functions \citep{lacey,sheth02}; the size distribution is
\bqa
m \left. \frac{dn}{dm} \right|_{\rm box} & = & \sqrt{\frac{2}{\pi}} \
\frac{\bar{\rho}}{m} \ \left| 
  \frac{d \ln \sigma}{d \ln m} \right| \
  \frac{\sigma^2(m)}{[\sigma^2(m) - \sigma^2_{\rm box}]^{3/2}}
  \nonumber \\
& &  \times B_0 \exp \left[ - \frac{B^2(M,z)}{2 [\sigma^2(m) -
        \sigma^2_{\rm box}]} \right].
\label{eq:dndmbox}
\eqa
The results are compared to the ``true'' size distribution in the
bottom panel of Figure \ref{fig:dndr-comp}, assuming $L_{\rm box}=
10h^{-1} \Mpc = 14.3 \Mpc$; this is the same size as the simulations
discussed by \citet{sokasian03a}, which are among the largest volumes
that have been used in numerical studies of reionization.  We see that
even state of the art simulations probably underestimate the sizes of
ionized regions, especially during the late stages of overlap.
Studying these epochs will require box sizes $\ga 100 \Mpc$ for a fair
sample.

Before proceeding to discuss the statistical properties of this
distribution in more detail, we pause to note several caveats about
our model.  First, we do not necessarily have $\Qb = \zeta f_{\rm
coll}$, and hence our normalization may be incorrect.  One reason is
that equation (\ref{eq:fcoll}) breaks down on sufficiently small
scales.  Also, we use the linear fit $B(m,z)$ rather than the true
$\delta_x(m,z)$.  We therefore expect that our model will not properly
normalize the size distribution when the characteristic bubble size is
small.  Figure \ref{fig:qbar} indicates that this is indeed the case;
the dashed curves (which give the true ionized fraction) lie
significantly above the solid lines of our model at early times (when
the bubbles are small).  To solve this problem, we multiply our number
densities by $\zeta f_{\rm coll}/\Qb$.  Fortunately, the figure also
shows that this correction becomes negligible when $\Qb \ga 0.1$.
Thus the correction is only necessary when the bubbles have a small
effect anyway.  The required renormalization does increase if $m_{\rm
min}$ increases (because equation [\ref{eq:fcoll}] breaks down sooner)
but is never more than $\sim 20\%$ in the results we show.

Another caveat is that the excursion set formalism is fundamentally
Lagrangian: it follows \emph{mass elements} rather than \emph{volume
elements}.  More massive regions will have expanded less than their
underdense neighbors.  We do not include this difference in
calculating the radial sizes of the regions; this is reasonable
because regions that cross our barriers are still far from turnaround
at these early times.  Moreover, we of course assume spherical
\ion{H}{2} regions: in reality they have much more complex
shapes.

Finally, we emphasize several simplifications we have made to the
physics of reionization.  First, equation (\ref{eq:zeta}) neglects the
environmental dependence of the ionizing efficiency.  The
recombination rate, for example, increases with the mean local
density, making $\zeta$ a function of the density.  It is possible to
formulate a model of reionization based on the variation of the
recombination rate with density \citep{miralda00}; in such a model,
low density voids are ionized first.  However, the simulations show
instead that the densest regions tend to be ionized first, at
least if the sources are relatively faint and numerous
\citep{sokasian03a,ciardi03-sim}, in accord with our model.  The basic
reason is that high-redshift galaxies are strongly biased, so their
abundance increases faster than the local density.  Thus \ion{H}{2}
regions first appear in the densest regions and only then escape into
the voids.  Because subsequent generations of sources also form
primarily in the dense regions, this structure is preserved throughout
the ionization process.  We compare our model to one in which voids
are ionized first in \citet{furl04b}.

Recombinations will nevertheless be important in the detailed
ionization pattern inside of each ionized region.  Our model assumes
that such regions are \emph{fully} ionized; in reality, they will have
a complex ionization pattern determined by recombinations in clumpy
regions, ``shadowing'' of ionizing sources by high-column density
objects, and the sheets and filaments of the cosmic web.  In what 
follows, we ignore these complications, because they will imprint
signatures into e.g. the angular power spectrum on scales much
smaller than those characteristic of the bubbles as a whole.

\section{Constructing the Power Spectrum}
\label{ps}

We will now consider how to construct observable statistical measures
of the size distribution of \ion{H}{2} regions.  
If high signal-to-noise measurements are
possible, then of course $dn/dm$ of the ionized regions can be
measured directly.  We will instead consider regimes in which the
\ion{H}{2} regions cannot be mapped in detail but in which the
fluctuation pattern can still be measured statistically.  We will take
as an example the 21 cm power spectrum (ZFH04).  The fundamental
observable in 21 cm measurements is the brightness temperature of the
IGM \citep{field59a,mmr},
\bqa
\delta T_b \, & \approx \, 23 \,\, \psi & \times
\,\, \left( \frac{T_S - T_{\rm
CMB}}{T_S} \right) \left( \frac{\Omega_b h^2}{0.02} \right) \nonumber \\
& & \times \left[ \left(\frac{0.15}{\Omega_m h^2} \right) \, \left(
\frac{1+z}{10} \right) \right]^{1/2} \mkel.
\label{eq:dtb}
\eqa 
Here, $T_S$ is the hydrogen spin temperature and $\psi=\xh(1+\delta)$,
where $\xh$ is the local neutral fraction and $\delta$ is the local
overdensity; the other symbols have their usual meanings.  We will
assume throughout that $T_S \gg T_{\rm CMB}$, so that $\delta T_b$ is
independent of the spin temperature; this is expected to be a good
assumption soon after ionizing sources turn on (see ZFH04 for a more
detailed discussion).

The simplest statistical description of the 21 cm field is provided by
the power spectrum of $\psi$.  In this section we will describe how to
construct this function.  We will find it simpler to work in terms of
the correlation function $\xi_\psi$, which can be constructed
explicitly from its component fields (ZFH04):
\bq
\xi_\psi = \xixx ( 1 + \xidd) + \bxh^2
\xidd + \xi_{x\delta}( 2 \bxh + \xi_{x\delta}).
\label{eq:xipsi}
\eq 
Here $\xixx$ is the correlation function of $\xh$, $\xi_{\delta
\delta}$ is the correlation function of the density field, and $\xixd$
is the cross-correlation between the two fields.  The power spectrum
$P_\psi$ is then the Fourier transform of this quantity.  First, note
that $\xidd$ can be obtained directly using standard techniques such
as the halo model (e.g., \citealt{cooray02}).  This includes nonlinear
corrections to the power spectrum , although these are generally not
important on the scales of interest.  Because the $\xh$ field is
actually composed of isolated ionized regions, we find it more
intuitive to work in terms of the correlation function of the ionized
fraction, $x_i$.  However, note that by linearity the two correlation
functions are identical.  In the following sections we describe how to
build our model for $\xixx$.  We first describe the necessary limits
of the correlation function in \S \ref{limits} and explain why the
simplest models fail.  We go on to show how to compute $\xixx$ for a
randomly distributed set of \ion{H}{2} regions in \S \ref{model-xixx}
and then how to add source clustering in \S \ref{clus}.  Finally, we
compute $\xixd$ in \S \ref{xixd}.  Note that we will neglect redshift
space distortions and Jeans smoothing of the baryonic fluctuations.

\vspace{1cm}

\subsection{Restrictions on $\xixx$}
\label{limits}

We expect the joint probability distribution to have the form (ZFH04)
\bq
\langle x_1 x_2 \rangle \equiv \langle x_i(\br_1) x_i(\br_2) \rangle =
\bxi^2 + (\bxi - \bxi^2) f(r_{12}/R),
\label{eq:xxform}
\eq
with $r_{12}=|\br_1 - \br_2|$ and $R$ some characteristic bubble size.
The function $f$ has the limits $f \rightarrow 0$ for $r_{12} \gg R$
and $f \rightarrow 1$ as $r_{12} \rightarrow 0$.  That is, if two
points are separated by a distance much smaller than the size of a
typical \ion{H}{2} region they will either both be ionized by the same
bubble, with probability $\bar{x}_i$, or both neutral.  But if they are
well-separated relative to the bubble size they must reside in
distinct \ion{H}{2} regions, and the probability approaches $\bar{x}_i^2$.
The correlation function is then $\xixx=\langle x_1 x_2
\rangle - \bar{x}_i^2$.

In addition, we must note that $x_i$ can take values only in a
restricted range from zero to unity.  This means that $\bxi=1$ implies
$x_i=1$ \emph{everywhere}, so that the correlations vanish.  Thus, we
need $\xixx =0$ when both $\bxi=0$ and $\bxi=1$.  We note that this
limit is not satisfied by many simple models.  For example,
\citet{santos03} argue that $\xixx \propto x_i^2 \xidd$ times some
filter function that damps fluctuations inside the bubbles; this form
does not have the correct limiting behavior.  \citet{santos03} instead
enforce the limit by choosing the bubble size to approach infinity at
overlap (however, it does not have the correct behavior on large
scales when $x_i < 1$).  We require a prescription that
self-consistently satisfies the above limit, independent of our
prescription for bubble sizes.  For example, we wish to contrast the
model described in \S \ref{bubbles}, in which the individual bubble
sizes approach infinity at overlap, with other models in which overlap
is achieved by increasing the number density of bubbles but not their
sizes (such as placing \ion{H}{2} regions around each galaxy; see
Figure \ref{fig:dndr-comp}).

\subsection{A Model for $\xixx$}
\label{model-xixx}

Assuming that the sources are uncorrelated, the probability of having
$N$ sources in a volume $V$ follows a Poisson distribution.  Thus the
probability that at least one source falls into this region is $1 -
e^{-n V}$, where $n$ is the number density of sources.  We begin by
assuming that all the bubbles have the same size $V_I$.  The proper
definition of the ionized fraction is
\bq
\bxi = 1 - e^{-n V_I} = 1 - e^{-\Qb}
\label{eq:xion}
\eq
We let $V_o(r)$ be the volume of the overlap
region between two ionized regions centered a distance $r$ apart.  The
probability that both $\br_1$ and $\br_2$ are ionized has two parts.
One possibility is that a single source can ionize both points if it
is inside $V_o(r_{12})$.  Alternatively, two separate sources can
ionize the two points.  The total joint probability is then 
\bq
\langle x_1 x_2 \rangle = (1 - e^{-n V_o}) + e^{-n V_o}  [1 - e^{-n
    (V_I - V_o)}]^2, 
\label{eq:x1x2poiss}
\eq
where $V_o \equiv V_o(r_{12})$.  One can easily verify that the
formula obeys all of the restrictions described in \S \ref{limits}.

We now wish to relax the assumption of a single bubble size; we will
index the bubble size with $m$ for reasons that are probably obvious.
We first calculate the probability that a given point is ionized.
Around that point, there is a sphere of radius $R(m_1)$ where even the
smallest allowed source (with index $m_1$) is able to ionize the
point.  Let $P_1$ be the probability that this occurs.  If that does
not happen, there is a shell surrounding this sphere in which a source
of radius $R(m_2)$ can ionize our point; let $P_2$ be the probability
of there being a source of large enough size in this region.  If this
does not happen, there is another shell in which a source of radius
$R(m_3)$ can ionize the point, etc.  We can write the total
probability $\bxi$ that the point is ionized as 
\bqa
\bxi & = & P_1 + (1-P_1)P_2 + (1-P_1)(1-P_2)P_3 + ... \nonumber \\
& = & 1 - \exp \left[\sum_m \ln(1-P_m) \right] \nonumber \\
& = & 1 - \exp \left[\sum_m n(m) V(m) \right],
\label{eq:pclus}
\eqa
where $n(m)$ is the number density of sources with index $m$.

The correlation function can be calculated in the same way as above.
One possibility is that the two points are ionized by the same source;
this requires at least one source of size $m$ to be in the appropriate
overlap region $V_o(m)$.  The probability for this to occur is
\bq
P(1 \ {\rm source}) = 1 - \exp \left[\sum_m n(m) V_o(m) \right].
\label{eq:psinglesource}
\eq
The other possibility is that the points are ionized by different
sources.  Again there is a succession of regions, this time of size
$V(m)-V_o(m)$, where a source of size $m$ can ionize one point but not
the other.  Thus [c.f. equation (\ref{eq:x1x2poiss})]
\bqa
\langle x_1 x_2 \rangle & = & \left\{ 1 - \exp \left( -\int dm
  \, \frac{dn}{dm} \, V_o(m) \right) \right\} \nonumber \\
& & + \exp \left( -\int dm \, \frac{dn}{dm} \, V_o(m) \right) \nonumber \\ 
& & \times \left\{ 1 -
  \exp \left( - \int dm \, \frac{dn}{dm} \, [ V(m) - V_o(m)]
  \right) \right\}^2,
\label{eq:x1x2cont}
\eqa
where we have taken the continuum limit for $m$ and now make the explicit
connection with the bubble mass.  For clarity of presentation, we have
suppressed the $r_{12}$ dependence of $V_o(m)$.  This function can be
written explicitly as 
\bq
V_o(m,r_{12}) = \left\{
\begin{array}{ll}
4\pi R^3(m)/3 - \pi r_{12} [R^2(m) - r_{12}^2/12], & r_{12}<2R(m) \\
0, & r_{12}>2R(m).
\end{array}
\right.
\label{eq:voverlap}
\eq
Here $R(m)$ is the radius of an ionized region of mass $m$.

Equation (\ref{eq:x1x2cont}) manifestly obeys all of the limits that
we expect, and in particular $\xixx$ is identically zero when
$\bxi=1$.  However, it does not treat the overlap of \ion{H}{2}
regions correctly.  We expect that when two ionized regions overlap,
they expand so as to conserve volume (because the number of ionizing
photons has not changed).  The above prescription does \emph{not}
conserve volume.  To solve this problem, we multiply the number
density by $[-\ln(1-\Qb)/\Qb]$.  We thus force $\bar{x}_i = \zeta
f_{\rm coll}$.  Note that, provided we assign the bubble size
distribution in such a manner that already includes overlap -- such as
that of equation (\ref{eq:dndm}) -- we would ideally like to avoid
\emph{any} overlap when constructing the power spectrum.  There is no
clean way to avoid this problem, however, and our solution is a
reasonable one.

\subsection{Clustering}
\label{clus}

The treatment of \S \ref{model-xixx} assumes that the ionized regions
are uncorrelated on large scales.  We can see this in equation
(\ref{eq:x1x2cont}) by noting that, when $V_{o}(m) \approx 0$,
$\langle x_1 x_2 \rangle \rightarrow \bxi^2$.  Thus the correlation
function vanishes on scales larger than the bubble size.  We now show
how to include (approximately) the correlations between the bubbles
induced by density modulations.  Unfortunately, the precise clustering
properties are difficult to determine because we only compute the
probability that a point is ionized by \emph{any} source; the root of
the difficulty is that multiple sources can ionize a single point in
our development, and it is not clear how to treat this possibility.
However, the following simplification suffices for our purposes.
We replace one of the factors in the last term of equation
(\ref{eq:x1x2cont}) by 
\bq 
\left\{ 1 - \exp \left( - \int dm \, \frac{dn}{dm} \, [ V - V_o]
 [1 + \bar{\xi}(r_{12},m)] \right) \right\}.
\label{eq:cluster}
\eq
Here, $\bar{\xi}(r_{12},m)$ is the excess probability that the point
$\br_2$ is ionized by a source of size $m$ given that $\br_1$ is
ionized.  Thus $\bar{\xi}$ is implicitly an average over all the
sources able to ionize the first point.  It is also approximate in
that we have neglected the variation of the correlation function
across $V(m)$.  We are thus assuming that such variation is small, or
that $r \gg R$.  Fortunately, this is precisely the regime we are
interested in:  for small $r$, $[V(m) - V_o(m,r)] \rightarrow 0$
and the correlations are strongly suppressed anyway.  This is a
manifestation of the fact that we care only if a point is ionized at
all, not whether it is ionized by a single or many sources.

We now assume that the correlation function can be
written 
\bq
\bar{\xi}(r,m) = \bar{b} b(m) \xidd(r),
\label{eq:xibar}
\eq
where $\xidd(r)$ is the correlation function of the dark matter,
$b(m)$ is the linear bias of a source of mass $m$, and 
\bq
\bar{b} = \Qb^{-1} \int dm \, b(m) V(m) \frac{dn}{dm}
\label{eq:bbar}
\eq
is the bias averaged over the bubble filling factor.  The weighting by
$V$ enters because the probability that our point is ionized by an
object of size $m$ is proportional to the object's volume.  Finally,
\citet{sheth02} have shown that the bias of halos whose mass function
is fixed by a linear barrier is
\bq
b(m,z) = 1 + \frac{B_0(z)^2}{\sigma^2(m) B(m,z)}.
\label{eq:bias}
\eq
For $\bar{x}_i \la 0.5$, the bias factor is of order unity.  However,
it can become quite large near overlap.  This is because the
characteristic bubble size is fixed by $B \sim B_0 \sim \sigma$; thus
the second term is $\sim 1/B(m,z)$ which is large near overlap.  (At
overlap, $B \approx 0$ so that the entire universe is ionized.)
However, the correlations between bubbles in this regime turn out not
to be significant.  The typical correlation length at high redshifts
is $\la 1 \Mpc$; once the characteristic bubble size exceeds this
level the correlations can be neglected on the scales of interest (see
\S \ref{res-corr} below).

\subsection{Cross-Correlation with Density}
\label{xixd}

The final step is to compute the cross-correlation between the density
and ionization fields.  If we use the halo model for the density
field, this is straightforward:
\begin{eqnarray}
\langle x_i(\br_1) \delta(\br_2) \rangle & = & - \bar{x}_i + \int dm_h \,
\frac{m_h}{\bar{\rho}} \, 
n_h(m_h) \int d^3 \br_h \, u(\br_2 - \br_h|m_h) \nonumber \\
& & \times \left\{ 1 - \exp \left[ - \int dm_s \, \frac{dn}{dm_s}
  \nonumber \right. \right. \\ 
& & \times \left. \left. \int_{V(m_s,\br_1)} d^3  \br_s \, (1 + \xi_{m_s,m_h}(\br_h - \br_s)
  ) \right] \right\},
\label{eq:xd}
\end{eqnarray}
where $n_h(m_h)$ is the halo mass function, $\xi_{m_s,m_h}$ is the
excess probability of having a bubble of size $m_s$ given a halo of
mass $m_h$ at the specified separation, and $u({\bf x}|m_h)$ is the
normalized halo mass profile: $\int d^3 {\bf x} u({\bf x}|m_h) = 1$
\citep{cooray02}.

As written, this is a quadruple integral that is difficult to evaluate
numerically.  Fortunately, we can make several straightforward
simplifications.  First, we again write the correlation function as a
multiple of the dark matter correlation: $\xi_{m_s,m_h} \approx b(m_s)
b_h(m_h) \xidd$, where $b_h(m_h)$ is the usual halo bias \citep{mo96}.
Second, we note that halos at high $z$ are much smaller than both the
bubble size and the separations that we are interested in.  Thus we can
set $\br_h \approx \br_2$ in the exponential.  The integration over
the mass profile is then separable (and equal to unity).  Finally, we
note that with this approximation the average of the correlation
coefficient (i.e., the integral over $\br_s$) is also separable.  We
define
\bq
\langle \xi \rangle = \int dm \, \frac{dn}{dm} \, b(m) 
\int_{V(m,\br_1)} d^3\br_s \, \xidd(|\br_s - \br_2|), 
\label{eq:xieff}
\eq
where the integral over $d^3 \br_s$ is centered on the point $\br_1$.
The limiting behavior of $\langle \xi \rangle$ is easy to understand.
If the separation is much smaller than the characteristic bubble size
$R$, the integral will be dominated by sources at typical separations
$R$ and $\langle \xi \rangle \sim \bar{b} \bar{Q} \xidd(R)$; if on the
other hand $r \gg R$, the integral is dominated by sources at
separations $r\approx r_{12}$ and $\langle \xi \rangle \sim \bar{b}
\bar{Q} \xi_{\delta \delta}(r)$.  

With these two simplifications, we can rewrite equation (\ref{eq:xd})
as 
\bq
\langle x_1 \delta_2 \rangle = (1 -\bar{x}_i) \left\{ 1 -
\int dm_h \, 
\frac{m_h}{\bar{\rho}} \, n_h(m_h) \, e^{-b_h(m_h)\langle \xi
  \rangle} \right\}.
\label{eq:xibarapprox}
\eq
The limiting behavior of the cross-correlation is also easy to
compute.  We find that on large scales
\bq
\langle x_i(\br_1) \delta (\br_2) \rangle \approx (1 - \bxi) \bar{b}_h
\bar{b} \Qb \xidd(r_{12}), \qquad r_{12} \gg R
\label{eq:xixdlim}
\eq
where $\bar{b}_h$ is the mean halo bias (appropriately weighted) and
$R$ is the typical bubble size.  We see that $x_i$ and the density are
positively correlated on large scales, with a constant correlation
coefficient.  This happens because we associate ionized regions with
overdensities.  On small scales,
\bq
\langle x_i(\br_1) \delta (\br_2) \rangle \approx (1 - \bxi) [1 -
  e^{-\bar{b}_h \bar{b} \Qb \xidd(R)}], \qquad r_{12} \ll R.
\label{eq:xixdlim0}
\eq
Thus for small separations the two fields are essentially uncorrelated so
long as $\xidd(R)$ is not large; this is because the entire bubble is
ionized in our model, regardless of its internal density structure.

Finally, note that in this section we have computed the
cross-correlation between the ionized fraction and the density field.
However, in equation (\ref{eq:xipsi}) we want the cross-correlation
between the \emph{neutral} fraction and the density field.  This is
obviously $\xixd = - \langle x_i(\br_1) \delta (\br_2) \rangle$, so
$\xh$ and the density are anticorrelated.

\section{Results}
\label{res}

\subsection{The Correlation Function}
\label{res-corr}

\begin{figure}
\plotone{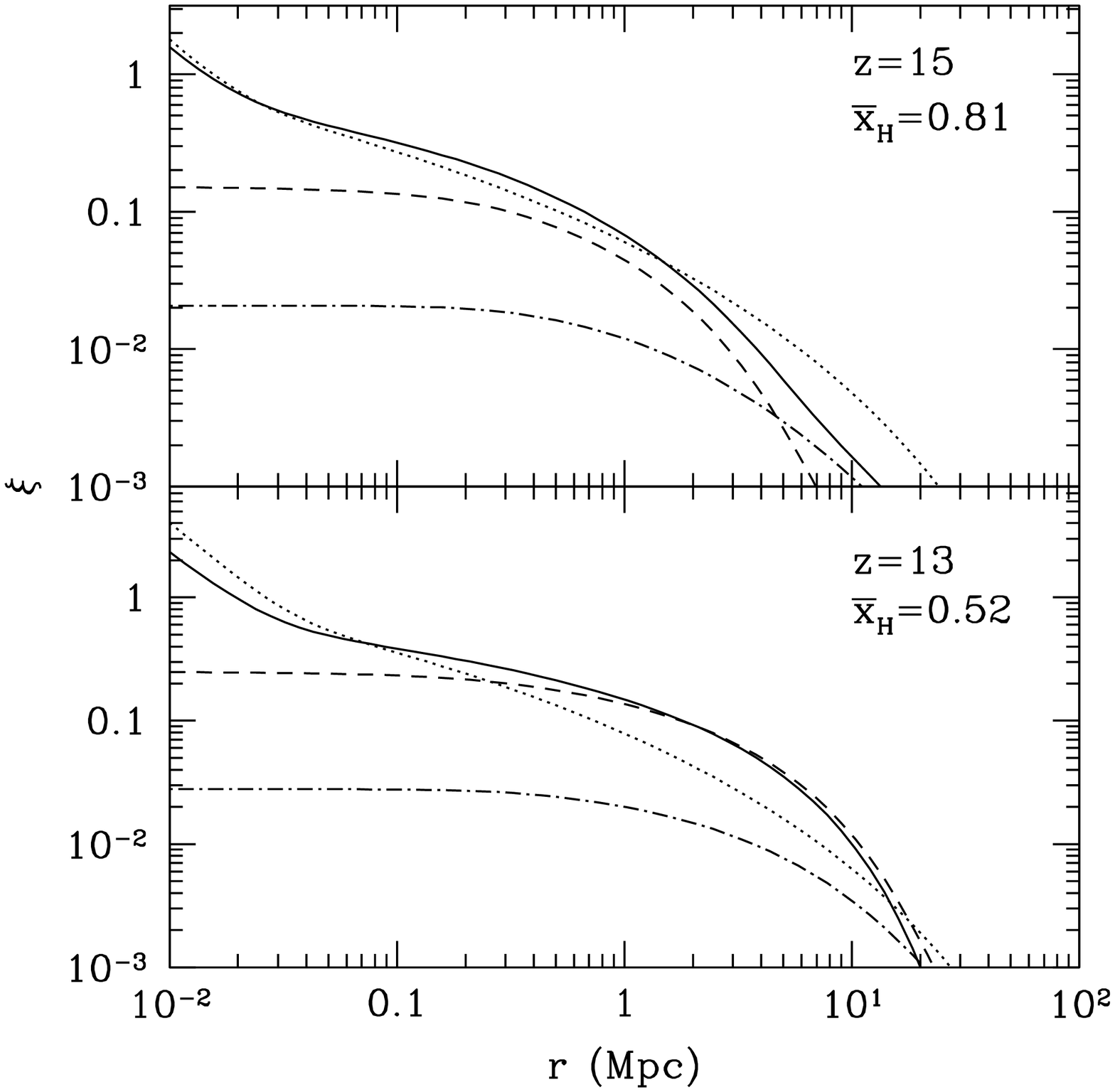}
\caption{The correlation functions at $z=15$ (top panel) and $z=13$
(bottom panel), assuming that $\zeta=40$.  In each panel, the solid,
dashed, dotted, and dot-dashed curves show $\xi_\psi$, $\xixx$,
$\xidd$, and $|\xixd|$, respectively.  (Note that the $\xh$ and
density fields are anticorrelated, so $\xixd<0$.)}
\label{fig:corr}
\end{figure}

We begin by describing the salient features of the correlation
function, $\xi_\psi$.  Figure \ref{fig:corr} shows example correlation
functions of density, $\xh$, and $\psi$ at $z=15$ (top panel) and
$z=13$ (bottom panel).  Both assume $\zeta=40$, which yields
$\bxh=0.81$ and $0.52$ at these two redshifts.  We also show the
absolute value of the cross-correlation $\xixd$.  First, we see that
as $r$ decreases from infinity, $\xidd$ first becomes shallow and then
steepens again at small separations.  The steepening occurs when the
one-halo term becomes important; generally, this is on scales smaller
than we will be able to probe with the next set of low-frequency radio
telescopes.\footnote{The halo model correlation function in this
regime also depends on the choice of halo mass function; we use the
\citet{press} mass function.  The \citet{sheth99} mass function
changes the one-halo term by $\la 20\%$ here.}

We can also see that $\xixx$ has the form expected.  At small
separations, the correlation is dominated by whether the two points
are in a single bubble or not; in this regime $\xixx \rightarrow \bxh
(1-\bxh)$.  At large separations, the points must each be inside
different bubbles, so $\xixx \rightarrow 0$; actually the clustering
term begins to dominate on sufficiently large scales and $\xixx
\propto \xidd$ (this is just visible in the top panel).  The
transition between these two regimes occurs sharply, on scales that
are comparable to the typical size of \ion{H}{2} regions.  In this
intermediate regime we can have $\xixx \gg \xidd$, provided that the
bubbles are sufficiently common and that the typical bubble size is
significantly larger than the correlation length of the density field.
(If not, $\xixx$ simply traces the correlation function of the density
field as it falls.)  The cross-correlation has similar behavior,
approaching a constant value at small separations and a constant
multiple of $\xixx$ on large scales.  Note that the cross-correlation
is negligible on small scales, essentially because our bubbles have no
internal structure, but it can be important on large scales.

The solid lines in Figure \ref{fig:corr} are $\xi_\psi$.  On
sufficiently small scales the dominant term in equation
(\ref{eq:xipsi}) is $\xixx \xidd$, and the correlation function is
simply proportional to the one-halo term of the density field.  On
extremely large scales, $\xi_\psi$ falls below $\xidd$ because of the
anti-correlation between density and $\xh$.  On intermediate scales,
the ionized regions add only a small perturbation to $\xi_\psi$ when
the neutral fraction is large, but dominate by a large factor in the
later stages of overlap.  This ``hump'' is the observational signature
that we seek.

\begin{figure}
\plotone{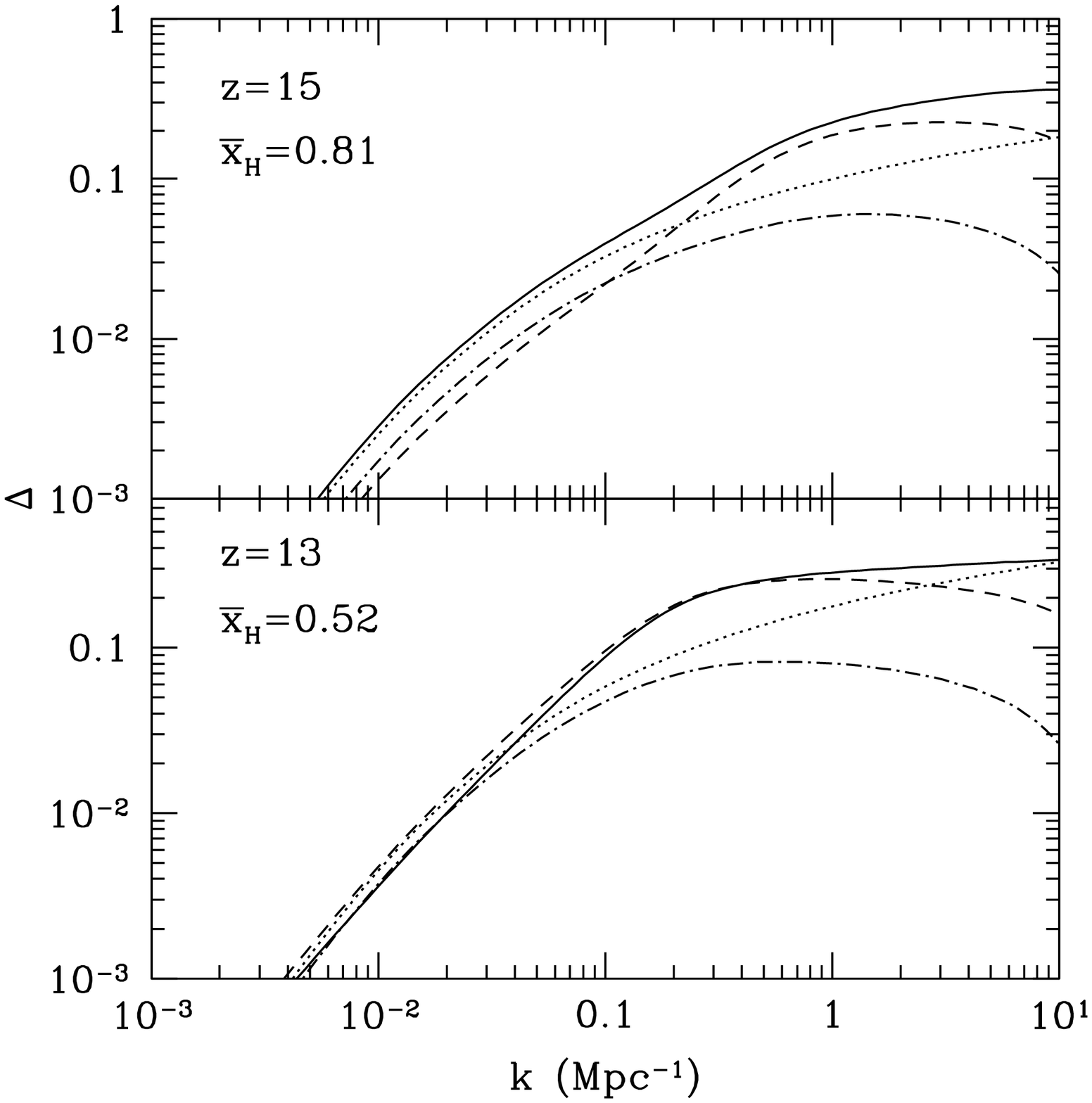}
\caption{The power spectra at $z=15$ (top panel) and $z=13$ (bottom
panel), assuming that $\zeta=40$.  In each panel, the solid, dashed,
dotted, and dot-dashed curves show $P_\psi$, $\Pxx$, $\xh^2 \Pdd$, and
$|\Pxd|$, respectively.  (Note that the $\xh$ and density fields are
anticorrelated, so $\Pxd<0$.)}
\label{fig:psexample}
\end{figure}

\subsection{The Power Spectrum}
\label{psexample}

Figure \ref{fig:psexample} shows the power spectra of the correlation
functions in Figure \ref{fig:corr}.  Here $\Delta^2(k)=(k^3/2\pi^2)
P(k)$ is the dimensionless power spectrum; we show the square root of
this quantity because in 21 cm studies $\delta T \propto \psi$ rather
than $\psi^2$.  The behavior is essentially as expected from the
discussion of the correlation functions above.  When the neutral
fraction is large, $P_\psi$ traces the density power spectrum, with
the ionized regions adding only a small perturbation on the relevant
scales.  On large scales, $P_\psi$ also traces the density power
spectrum, but $P_\psi < \Pdd$ because the bubbles are anti-correlated
with density.  Once overlap has become significant, $\Pxx$ dominates
the power spectrum.  In the bottom panels, the bubbles increase the
power on and above their characteristic scale by nearly an order of
magnitude.  On large scales, our model predicts that $P_\psi$ stops
tracing $\Pdd$ when $\bxh$ is small.  This is because the
characteristic bubble size approaches infinity near overlap.  The
number density of the bubbles thus decreases rapidly when $\bxh$ is
small, so the ``shot noise" becomes significant and dominates the
large scale power.  In this regime, $\Delta^2_\psi \propto k^3$, which
is approximately satisfied at $k \la 0.1 \Mpc^{-1}$ in the bottom
panel of Figure \ref{fig:psexample}.

This Poisson noise component is to some extent an artifact of the
simplifying assumptions inherent to our model.  We assumed that each
point is either fully ionized or fully neutral, neglecting effects
like recombination, clumpiness, shadowing, asphericity, etc.  Once the
\ion{H}{2} regions become sufficiently large, these effects will
determine the internal structure of the bubbles.  By effectively
decreasing the size and increasing number density of ionized regions,
they would therefore reduce the power on large scales, bringing
$P_\psi$ closer to the density power spectrum.  Including these
detailed effects will require numerical simulations.

We note that the power spectra shown here differ from those of ZFH04
in that the ``bubble" feature is significantly less pronounced.  This
is simply because that paper employed a ``toy" model in which the
bubbles had a single effective size at each redshift.  Also, our
models have relatively more power on large scales late in
reionization, because the ZFH04 analysis reduced the bubble size at these
stages while keeping the number density fixed.  Because the shot noise
decreases with the number density, the large scale power was small in
that case.

\begin{figure}
\plotone{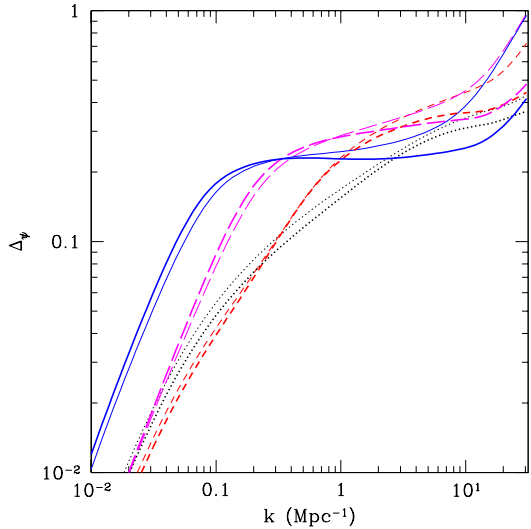}
\caption{The redshift evolution of $\Delta_\psi$ in the
  $\zeta=12$ (thin lines) and $\zeta=40$ (thick lines) models.  The
  curves are: $\bxh=0.96$ (dotted), $\bxh=0.8$ (short-dashed),
  $\bxh=0.5$ (long-dashed), and $\bxh=0.26$ (solid). The redshifts in
  the two models differ.}
\label{fig:psi12i40}
\end{figure}

\subsection{Reionization Histories}
\label{hist}

We now consider how the power spectrum evolves with time for some
simple reionization histories.  We will examine only single
reionization episodes here, although our formalism can accommodate
other, more complicated histories as well (see \citealt{furl04b}).
For reference, we note that a ``typical" Population II star formation
history, with $f_{\rm esc}=0.2$, $f_\star=0.05$, $N_{\gamma/b}=3200$,
and $n_{\rm rec}=3$, would have $\zeta\approx12$.  This gives a
reionization history similar to that of \citet{sokasian03a},
who adopted a Population II star formation rate based on
numerical simulations \citep{hs03,sh03}.
Population III stars can have $\zeta$ an order of magnitude larger
without much difficulty.  The evolution of $\bxh$ for some sample
histories is illustrated in Figure \ref{fig:qbar}.

\begin{figure}
\plotone{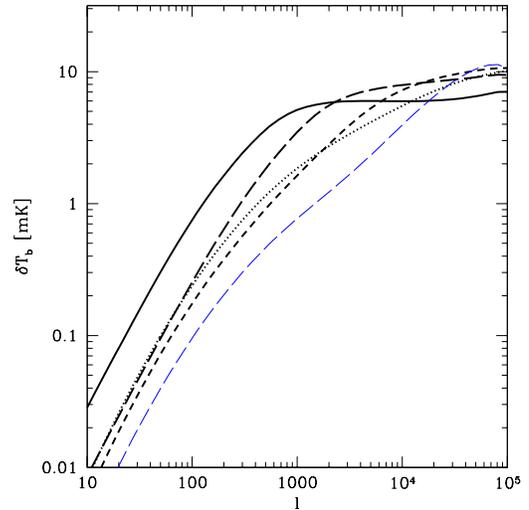}
\caption{The angular power spectrum of brightness fluctuations 
  in the $\zeta=40$ model.  The thick curves are for: $\bxh=0.96$
  (dotted), $\bxh=0.8$ (short-dashed), $\bxh=0.5$ (long-dashed), and
  $\bxh=0.26$ (solid).  The thin long-dashed curve is for the model
  where galaxies host their own individual HII regions ($\bxh=0.5$).}
\label{fig:cl}
\end{figure}

Figure \ref{fig:psi12i40} shows the evolution of the power spectrum
for $\zeta=40$ (thick lines) and $\zeta=12$ (thin lines).  The dotted,
short-dashed, long-dashed, and solid curves have
$\bxh=0.96,\,0.8,\,0.5$, and $0.25$, respectively.  These correspond
to $z\approx18,\,15,\,13$, and $12$ (for $\zeta=40$) and
$z\approx16,\,12,\,10$, and $9$ (for $\zeta=12$).  The most important
point is that the power spectrum evolves substantially throughout
reionization.  While $\bxh$ declines (and hence so does the mean
brightness temperature for 21 cm observations), the \ion{H}{2} regions
add power and more than compensate for the paucity of neutral gas.
Thus we confirm that statistical measurements can yield strong
constraints on the ionization history.  Surprisingly, the two curves
show very little difference on scales $k \la 10 \Mpc^{-1}$.  The
amplitudes, and especially the peak locations, closely match each
other.  On such scales the ionized regions dominate the power
spectrum, so this indicates that the bubble size distribution for a
fixed $\bxh$ and $m_{\rm min}$ is nearly independent of redshift and
$\zeta$.  In other words, the barrier $B(m,z)$ is determined primarily
by the neutral fraction.  This points to an interesting result of our
model: when scaled to the global neutral fraction, the morphology of
reionization does not vary strongly.  The invariance does break down
if $\zeta$ is extremely large (greater than several hundred) or if
$m_{\rm min}$ is large.  This is because the ionization field becomes
more sensitive to the highly-biased peaks in such cases.  It also
breaks at small scales, where $P_\psi \propto \Pdd$, because the
latter grows rapidly in this redshift range.  Finally, we
caution the reader that the behavior on large scales during the late
stages of reionization is not well-described by our model (see \S
\ref{psexample}).  Thus the trends we note may not hold in more
detailed models near overlap.

In Figure \ref{fig:cl} we show the angular power spectrum of the
brightness fluctuations, which measures the root mean square level of
fluctuations as a function of multipole $l$ (see ZFH04 for details).
To produce the figure we assumed that the observations were done with
perfect frequency resolution.\footnote{We have also neglected redshift
  space distortions; see ZFH04 for a discussion of their importance.}
In that case, the angular power spectrum traces $\Delta_\psi(k)$ with
the correspondence $l\sim k D$, where $D$ is the angular diameter
distance to the appropriate redshift.  As a result, all the features
in Figure \ref{fig:psi12i40} can be seen in Figure \ref{fig:cl}. For
an experiment with finite resolution the power spectra will be
somewhat different on small angular scales. An observed bandwidth
$\Delta \nu$ corresponds to a comoving distance
\bq
L \approx 1.7 \left( \frac{\Delta \nu}{0.1 \rm{Mhz}} \right) \left(
\frac{1+z}{10} \right)^{1/2} \left( \frac{\Omega_m h^2}{0.15}
\right)^{-1/2} \Mpc.
\label{eq:lcom}
\eq 
We denote by $l_{\Delta \nu}$ the multipole corresponding to the angle
subtended by $L$. The angular power spectrum will be given by
$\Delta_\psi(l/D)$ for $l<l_{\Delta \nu}$ but will be proportional to
$\Delta_\psi(l/D) \times (l_{\Delta \nu}/l)$ for $l>l_{\Delta
\nu}$. Thus, on small angular scales the shape of the spectrum will
change and the fluctuations will be reduced. For frequency resolutions
around $\Delta\nu \sim 0.2 \ {\rm MHz}$, realistic for upcoming
experiments, this change in behavior will happen at arcminute scales,
or equivalently $l\sim 10^4$ (see ZFH04 for details).

\begin{figure}
\plotone{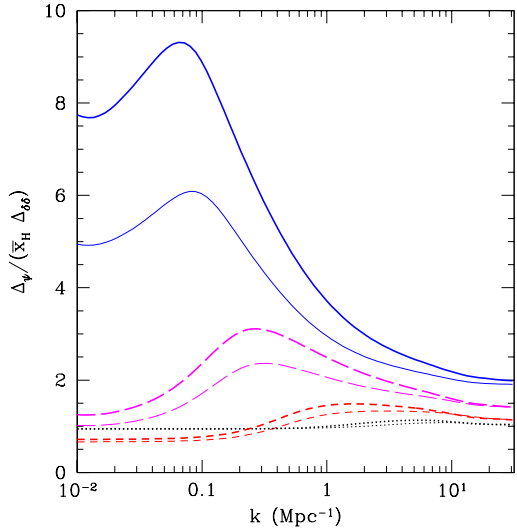}
\caption{The redshift evolution of $\Delta_\psi/(\bxh
  \Delta_{\delta\delta})$ in the $\zeta=12$ (thin lines) and
  $\zeta=40$ (thick lines) models.  The curves are: $\bxh=0.96$
  (dotted), $\bxh=0.8$ (short-dashed), $\bxh=0.5$ (long-dashed), and
  $\bxh=0.26$ (solid). The redshifts in the two models differ.}
\label{fig:ratioi12i40}
\end{figure}

The behavior on small scales is better understood through Figure
\ref{fig:ratioi12i40}, which shows the ratio between $\psi$ in our
model and that of a uniformly ionized IGM with the same $\bxh$.  At
large $k$, the curves each approach constant values nearly independent
of $\zeta$.  That is, $P_\psi$ is nearly proportional to $\Pdd$ at
large wavenumbers.  In fact, it is easy to see from equation
(\ref{eq:xipsi}) that $P_\psi/\Pdd \sim \bxh$.  (There is actually an
additional bias factor that affects the limiting value.)  Figure
\ref{fig:ratioi12i40} also shows explicitly that the bubble peak moves
to larger scales and higher amplitude as reionization progresses.  At
a fixed neutral fraction, the peak amplitude relative to the
underlying density field is largest at early times.  This is partly
because the large scale bubble characteristics are nearly independent
of redshift while the density power spectrum grows as $(1+z)^{-1}$ and
partly because the bias increases with redshift.  Thus (from a
theoretical standpoint) distinguishing between a uniformly ionized
medium and one with fully ionized regions will be somewhat easier if
reionization occurs earlier.

\begin{figure}
\plotone{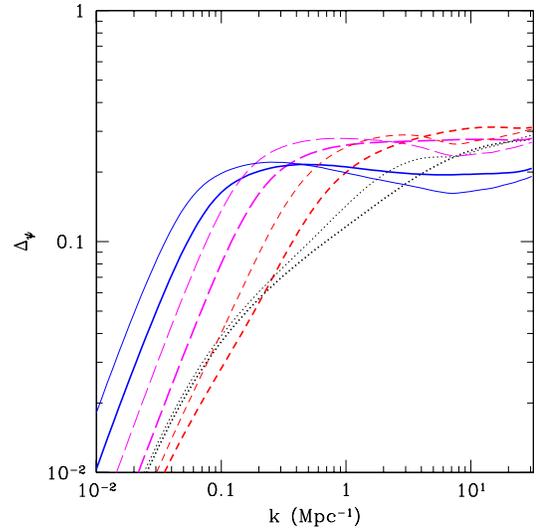}
\caption{The redshift evolution of $\Delta_\psi$ in the $\zeta=500$
  (thin lines) and $\zeta=90$ (thick lines) models.  In the latter
  case, we decrease $m_{\rm min}$ by an order of magnitude from the
  default model.  The curves are: $\bxh=0.96$ (dotted), $\bxh=0.8$
  (short-dashed), $\bxh=0.5$ (long-dashed), and $\bxh=0.24$ (solid).
  The redshifts in the two models differ slightly.}
\label{fig:psihiz}
\end{figure}

Figure \ref{fig:psihiz} compares two early reionization scenarios in
which overlap is complete by $z \sim 16$.  The thin lines assume
$\zeta=500$ and the usual value of $m_{\rm min}$; the thick lines have
$\zeta=90$ but decrease $m_{\rm min}$ by a factor of 10.  The
redshifts in the Figure are slightly different, ranging from
$z=22.5$--$17.2$ in the $\zeta=500$ model and $z=24$--$17$ in the
other.  As one might expect from the above discussion, there is
relatively little difference between the two histories, at least in
terms of the power spectra.  Our model predicts that features occur on
slightly larger angular scales for $\zeta=500$ and have slightly less
small-scale power.  The bubble feature also appears earlier in the
$\zeta=500$ model.  This is because the bubbles are initially
significantly larger in this case (both because the sources are
stronger and because they are more rare).  However, overall the
differences are small enough that distinguishing them will probably
require a more accurate treatment of reionization through numerical
simulations.  When compared to Figure \ref{fig:psi12i40}, the major
difference is the increased large scale power.  This too is because
the \ion{H}{2} regions have a smaller number density and hence the
shot noise term becomes significant earlier.

\subsection{\ion{H}{2} Regions Around Individual Galaxies}

\begin{figure}
\plotone{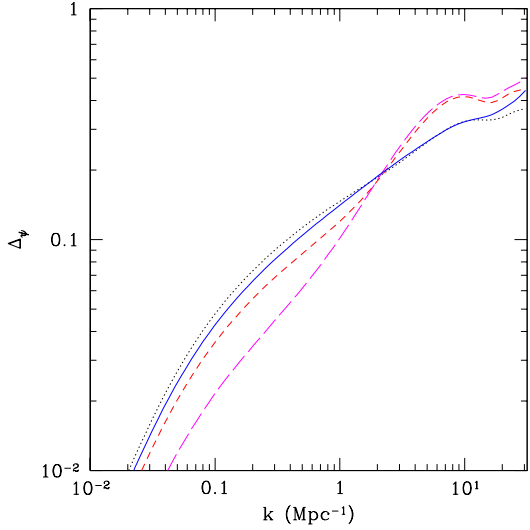}
\caption{The redshift evolution of $\Delta_\psi$ if we assume
  that each galaxy hosts its individual \ion{H}{2} region.  All curves
  assume $\zeta=40$.  The lines are: $\bxh=0.96$ (dotted), $\bxh=0.8$
  (short-dashed), $\bxh=0.5$ (long-dashed), and $\bxh=0.26$ (solid).}
\label{fig:psigal}
\end{figure}

The formalism described in \S \ref{ps} is general and does not rely on
our model for the size distribution of \ion{H}{2} regions.  We can
thus apply it to a scenario in which each galaxy hosts its own ionized
region; the size distribution is then shown in the top panel of Figure
\ref{fig:dndr-comp}.  Such a model would be relevant if galaxies were
truly randomly distributed.  Figure \ref{fig:psigal} shows the power
spectra for several redshifts, again assuming $\zeta=40$.  These
curves can therefore be compared directly to the thick lines in Figure
\ref{fig:psi12i40}.  The difference between the models is substantial
and it can also be seen in the angular power spectrum (see Figure
\ref{fig:cl}).  First, in the present case the peak not only occurs at
smaller scales, but its location does not shift significantly as
reionization is approached.  This is because overlap is achieved not
by increasing the sizes of individual galaxies (though that does
slowly happen) but instead by rapidly increasing their number density.
Second, the large scale behavior is significantly different: in the
present case $P_\psi \propto \Pdd$, and is always \emph{smaller} than
the power assuming a fully neutral medium.  This is because the number
of ionized regions is so large, making the shot noise negligible.  The
large scale power is minimized when $\bxh \approx 0.5$, because at
that point the (anti-correlated) density and $\xh$ fields nearly
cancel each other out.  
The long-dashed thin curve in Figure \ref{fig:cl} shows the
corresponding angular power spectrum for this model with $\bxh=0.5$.
We see that $\delta T_b$ is nearly featureless on scales $l\la10^4$,
with the bubbles appearing only at sub-arcminute scales.  There is
also a significant deficit of large-scale power.  This comparison
emphasizes the importance of considering large \ion{H}{2} regions when
describing reionization.

\section{The Pixel Distribution Function \& \\ Non-Gaussian Signatures}
\label{nongauss}

The power spectrum that we have computed in the last two sections is
only one statistical characterization of the maps.  Its widespread use
in many studies, and in particular for the CMB and galaxy redshift
surveys, stems from the fact that linear density perturbations are
nearly gaussian.  For such a field, the power spectrum completely
describes its statistical properties.  In our case, however, the
ionization pattern is not gaussian, so other characterizations may be
more powerful.  Here, we estimate the expected probability distribution
function (PDF) of the dimensionless pixel neutral hydrogen density
$\psi$ with an eye toward developing such alternative diagnostics.  We
caution the reader that this treatment will only be approximate; more
sophisticated techniques, and probably numerical simulations, will be
required for accurate descriptions.

We begin by selecting our pixel size and the effective enclosed mass
$\mpix$ (assuming the mean density for the volume-to-mass conversion).
Conceptually, we divide the \ion{H}{2} regions into those larger than
$\mpix$ and those smaller than this mass scale.  The PDF will have two
components: a delta function at $\psi=0$ made up of the large bubbles
and a broader distribution at $\psi>0$ composed of pixels with a mix
of neutral gas and small bubbles.  We can construct the second
component as follows.

Suppose we select a region with smoothed overdensity $\delta_0$.  We
first assume that the density field is gaussian; this is a reasonable
approximation on most of the scales of interest (see below).  Then we
know $f_{\rm coll}(\delta_0)$ and hence also $\psi(\delta_0)$ provided
that we include only sources contained inside this pixel.  If these
were the only relevant ionizing sources, we could construct the PDF of
$\psi$ in the usual way: $p(\psi) = p(\delta) |d\delta/d\psi|$.
However, we must also account for ionizing sources outside of our
selected region; in other words, we need the probability that a pixel
with a given $\delta_0$ is contained within one of the large bubbles.
In terms of the excursion set formalism, we wish to compute the
probability that a particular trajectory that passes through our pixel
scale at the appropriate density has passed above our barrier
$B(\sigma^2)$ for some $\sigma^2<\sigma^2_{\rm pix}$; i.e., that it
has been incorporated into a large-scale \ion{H}{2} region.  We let
$p(\delta^f_{\sigma}=B_{\sigma}|\delta_{\sigma_{\rm pix}}=\delta_0)$
be the conditional probability that a trajectory first crosses the
barrier at $\sigma^2$ given that it has $\delta(\sigma^2_{\rm
pix})=\delta_0$.  The total probability that our pixel is fully
ionized is then
\bq
p_i(\delta_0) = \int_0^{\sigma^2_{\rm pix}} d \sigma^2 \
p(\delta^f_{\sigma}=B_{\sigma}|\delta_{\sigma_{\rm
pix}}=\delta_0).
\label{eq:pi}
\eq

To compute the integrand, we use Bayes' Theorem:
\bqa
p(\delta^f_\sigma=B_\sigma|\delta_{\sigma_{\rm
pix}}=\delta_0) \ p(\delta_{\sigma_{\rm pix}}=\delta_0) = \nonumber \\
p(\delta_{\sigma_{\rm pix}}=\delta_0|\delta^f_{\sigma}=B_\sigma)
\ p(\delta^f_\sigma=B_\sigma).
\label{eq:bayes}
\eqa
Here $p(\delta_{\sigma_{\rm pix}} = \delta_0| \delta^f_\sigma =
B_\sigma)$ is the conditional probability that a trajectory has
$\delta_0$ on our pixel scale given that it first crosses the barrier
at $\sigma^2$. This, as well as $p(\delta)$, are gaussian, while
$p(\delta^f_\sigma=B_\sigma)$ is the first-crossing distribution
described in \S \ref{bubbles}.  Thus
\bqa
p(\delta^f_\sigma=B_\sigma|\delta_{\sigma_{\rm
pix}}=\delta_0) & = & \frac{B}{\sqrt{2 \pi}} \
  \frac{\sigma_{\rm pix}}{\sigma^3(\sigma_{\rm pix}^2 -
    \sigma^2)^{1/2}} \nonumber \\ & & \times \exp \left[ -
    \frac{(\delta \sigma^2 - B 
      \sigma^2_{\rm pix})^2} { 2 \sigma^2 \sigma^2_{\rm pix}
      (\sigma^2_{\rm pix} - \sigma^2)} \right].
\label{eq:pfb}
\eqa
If we instead used a constant barrier, note that $p_i$ could be
written explicitly using mirror symmetry about the barrier, as in
the standard derivation of the Press-Schechter mass function
\citep{bond91}. 

The PDF of $\psi$ is then simply
\bq
p(\psi) = \sum_{i=1}^2 p(\delta_i) \left| \frac{d \delta}{d \psi}
\right|_{\delta_i} [1-p_i(\delta_i)].
\label{eq:psipdf}
\eq
The summation occurs because $\psi(\delta)$ is not monotonic.  Clearly
$\psi$ approaches zero as the density decreases.  But it also
approaches zero as $\delta \rightarrow \infty$.  At high densities,
more ionizing sources reside in the pixel and $\xh$ approaches zero.
Thus, $\psi$ has a maximum value $\psi_{\rm max}$ at any pixel size and
redshift, and any given $\psi$ corresponds to two densities
$\delta_i$.  Also note that the PDF becomes singular (though still
integrable) at $\psi_{\rm max}$.  For analytic studies, it can thus be
easier to consider the cumulative distribution function, $P(>\psi)$,
which is always non-singular.

\begin{figure}
\plotone{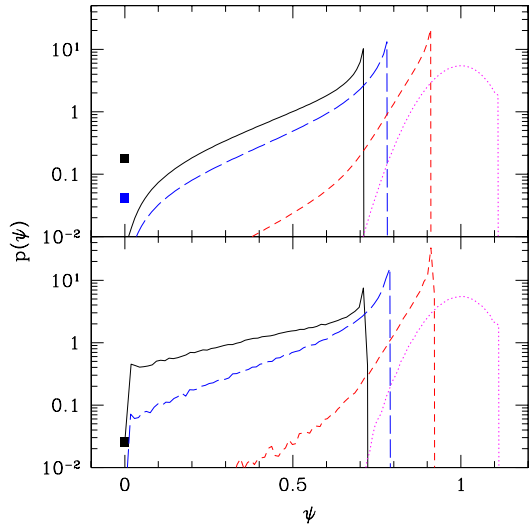}
\caption{The probability distribution $p(\psi)$ for $\mpix=10^{13}
  \msun$ and $\zeta=40$.  Dotted, short-dashed, long-dashed, and solid
  curves are for $z=20,\,16,\,14$, and 13, respectively.  The top
  panel shows the PDF of equation (\ref{eq:psipdf}), while the bottom
  panel approximately includes overlap with large bubbles (see text).
  The filled squares indicate the total probability of the pixel being
  fully ionized.  In the top panel, the upper and lower squares are
  $z=13$ and $z=14$, respectively; in the lower panel, only $z=13$ has
  a non-negligible probability. }
\label{fig:ppsi}
\end{figure}

Some examples are shown in Figure \ref{fig:ppsi}.  We choose
$\mpix=10^{13} \msun$, corresponding to $R \approx 3.9 \Mpc$ or an
angular diameter $\theta \approx 2.7'$ at $z=13$.  The top panel shows
$p(\psi)$ at $z=20,\,16,\,14$, and $13$ (dotted, short-dashed,
long-dashed, and solid curves, respectively) in the $\zeta=40$ model.
At $z=20$, the PDF resembles a truncated gaussian.  But even at
$z=16$, when $\bxh=0.88$, ionizations have completely transformed the
shape of the PDF.  As described above, this is because the higher
density regions have larger $f_{\rm coll}$ and hence smaller neutral
fractions.  The high tail of the density distribution is thus folded
over around $\psi_{\rm max}$ (which corresponds to $\delta \sim 0$).
Moreover, when the density is close to the mean value, $\xh$ is nearly
linear in the overdensity, so the peak can be quite sharp.  As the
bubbles grow, the peak shifts to smaller $\psi$ and the tail grows
rapidly.  This occurs because high-density pixels have a higher
probability of being mostly ionized by bubbles comparable to their own
size.  Once the characteristic bubble size passes $\mpix$ (which
happens just before $z=13$ for these parameters, see Figure
\ref{fig:dndr}), the probability of $\psi=0$ rises rapidly as well.

\begin{figure}
\plotone{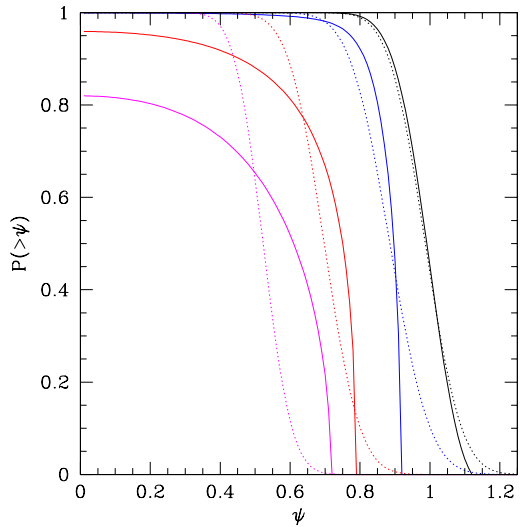}
\caption{The cumulative distribution function of $\psi$, assuming
  $\mpix=10^{13} \msun$ and $\zeta=40$.  The solid curves assume our
  model, while the dotted curves assume a uniformly ionized medium.
  In each case, the curves are for $z=20,\,16,\,14$, and 13, from top
  to bottom.}
\label{fig:psicdf}
\end{figure}

Another way to look at this is through $P(>\psi)$, which we show in
Figure \ref{fig:psicdf}.  The solid curves are for the above model,
while the dotted curves show the same function for a uniformly ionized
universe.  (In the latter case the PDF is just a gaussian.)  The
curves are for $z=20,\,16,\,14$, and $13$, from top to bottom.  We see
clearly that even when the ionized fraction is only a few percent, the
high-density tail is truncated and folded over to add to the
small-$\psi$ component.  It is obvious from this figure that the
variance is \emph{smaller} in the $f_{\rm coll}$ model during the
early phases of reionization; this is again because high density
regions and ionized bubbles are strongly correlated.  However, as the
bubbles grow $P(\psi)$ rapidly spreads toward $\psi=0$; this
corresponds to fluctuations in the ionized fraction dominating the
power spectrum.  This figure also explicitly shows that the
probability of having a fully ionized region is negligible until $z
\sim 13$.

This version of the PDF is not necessarily the natural choice for
comparing to observations, because we have assigned all of the large
bubbles to pixels with zero flux.  In a real survey, pixels could
intersect a fraction of a bubble.  In the limits in which the
characteristic bubble size is either much smaller or much larger than
the pixel size, this complication can be neglected.  We can estimate
the PDF in the intermediate range if we assume (i) that the fully
ionized part of a pixel is uncorrelated with the mean density of the
rest of the pixel, and (ii) that the part outside of a large bubble
follows the distribution of equation (\ref{eq:psipdf}); we thus
neglect the fact that this component is smaller than the full pixel
size.  In this case, the only additional component we need is the distribution
of the fraction $f_i$ of the pixel that is inside of large ionized bubbles.
This can be written
\bq 
p(f_i>f_0) = 1 - \exp \left[ - \int dm \, \frac{dn}{dm} V_{\rm
max}(m,f_0) \right],
\label{eq:}
\eq
where $V_{\rm max}(m,f_0)$ is the volume within which a bubble of the
specified mass will cover a fraction $f_0$ or larger of the pixel; it
is a straightforward generalization of equation (\ref{eq:voverlap}).

In this case we simulate $p(\psi)$ through a Monte Carlo algorithm.
We randomly select both $f_i$ and $\psi_n$ using their respective
cumulative distribution functions.  Here, $\psi_n$ refers to the part of
the pixel outside of the large bubble, so $\psi=f_i \psi_n$.  Some
examples of $p(\psi)$ are shown in the bottom panel of Figure
\ref{fig:ppsi}.  The effects of overlap with fully ionized regions are
obviously negligible when the pixel size is significantly larger than
the typical bubble size.  When the two scales are comparable at $z
\sim 13$, overlap spreads the delta function at $\psi=0$ out, but it
does not qualitatively affect the results.  If we had included
correlations, the effect of overlap would be slightly smaller than
shown here because pixels near $\psi_{\rm max}$ are less likely to be
near ionized regions than pixels that are already mostly ionized.

We have assumed throughout this section that the density field has a
gaussian distribution.  This is obviously only relevant on scales
where the power spectrum is approximately linear.  Our approximation
breaks down when higher moments of the density field become
significant.  As an example we consider the third moment of
$p(\delta)$.  The skewness due to pure gravitational instability is
$\langle \delta^3 \rangle = S_3 \langle \delta^2 \rangle^2$, where
$S_3 = [34/7 - (3+n)]$ if $P(k) \propto k^{-n}$
\citep{bouchet92,juszk93}.  The skewness only becomes significant at
$m \la 10^{12} \msun$ for $z \ga 10$, so our choice of $\mpix$ is
reasonable.

We thus see that the boosts in power from the \ion{H}{2} regions are
actually a result of a qualitative change in $p(\psi)$ and not simply
an increase in the variance of a gaussian.  This shows explicitly that
exploring statistics beyond the power spectrum is crucially important
to understanding the 21 cm signal.  From the estimates we have made
here, the optimal tests are not immediately obvious; we will defer
closer examination to the future.  Of course, our model is only
approximate, as we treat the perturbations in Lagrangian rather than
real space and neglect large-scale correlations and recombinations.
All of these complications will be important in determining the true
shape of the PDF and to what extent observations can distinguish
different models, but none will change our general conclusion about
the importance of non-gaussianity.  For example, the shape of the PDF
is determined entirely by our model for reionization; in other models
with comparable power the PDF could still have a qualitatively
different shape.  We explore one such example in \citet{furl04b}.

\section{Discussion}
\label{disc}

In this paper, we have described a new analytic model for the size
distribution of \ion{H}{2} regions during reionization.  While the
overlap process is usually described in terms of Str{\" o}mgren
spheres around individual galaxies, recent cosmological simulations of
reionization have demonstrated that the ionized regions are much larger than
naively expected, even early in the reionization process.  Here we
have shown that the size distribution can be understood in terms of
large-scale features in the density field, and we have constructed the
size distribution with an approach analogous to the standard
derivation of the \citet{press} mass function for collapsed halos.
The model has only two input parameters (if the cosmology is fixed):
the ionizing efficiency of collapsed objects and the minimum mass halo
that can host a luminous object.  Interestingly, at a fixed neutral
fraction the characteristic size of the bubbles is fairly insensitive
to these input parameters, suggesting that the morphology of
reionization is close to invariant (at least for simple, single
reionization episodes).  While we make several simplifying
approximations in the model (see the discussion in \S \ref{bubbles}),
it provides a self-consistent approach that reproduces the qualitative
features of simulations.  In the future, the model must be
quantitatively compared to simulations; however, an accurate
comparison requires simulations with large ($\sim 100^3 \Mpc^3$)
volumes because of the effects of the large-scale density field (see
Figure \ref{fig:dndr-comp}, as well as \citealt{barkana03}).

The morphology of reionization has implications for a number of
observables.  Most important, regardless of the technique, we predict
large \ion{H}{2} regions that should be feasible to detect, whether
through quasar absorption spectra \citep{miralda00,barkana02},
Ly$\alpha$ lines at extremely high-redshifts
\citep{pello04,loeb04,ricotti04,gnedin04,cen04,bart04}, or 21 cm
tomography, For example, the noise in a 21 cm map is proportional to
the square root of the bandwidth and, more important, to the square of
the pixel size.  Our model predicts substantial contrast on relatively
large scales of several arcminutes, which should make detections
easier.  Note, however, that in many next-generation experiments like
LOFAR the ``bandwidth" can be chosen \emph{after} the observations are
complete, so the expected scale of the bubbles need not determine the
experimental design \citep{morales03}.

Even if high signal-to-noise detections of individual \ion{H}{2}
regions are not available, we have shown that statistical measurements
of the size distribution can still strongly constrain reionization.
We first constructed the power spectrum of fluctuations in the neutral
density, including both density fluctuations and the ionized regions.
We found that the ionized regions imprint clear features on the power
spectrum and amplify the power by a factor of several during the
middle and late stages of reionization.  Most important, the power
spectrum evolves throughout reionization, allowing us to map the time
history of reionization.  For 21 cm observations, the large-scale \ion{H}{2}
regions predicted by our model put the features at $l \la 10^4$ (or
$\theta \ga 2\arcmin$). This matches well to the scales able to be
probed by upcoming experiments like PAST, LOFAR, and SKA (ZFH04).  If, on the
other hand, reionization occurred through the overlap of \ion{H}{2}
regions around individual galaxies, the features in the power spectrum
would appear at much smaller scales, perhaps beyond the reach of these
instruments.  We have also shown explicitly that the
ionized bubbles induce qualitative changes to the initially gaussian
neutral density distribution.  This suggests that statistical
measurements beyond the power spectrum can offer probes of the physics
of reionization that may be even more powerful.  

Our results also have important implications for other measurements.
For example, one of the most successful strategies for targeting
high-redshift galaxies is by searching for strong Ly$\alpha$ emitters.
If the galaxy is embedded in a mostly neutral medium, Ly$\alpha$
absorption from the IGM can have a substantial effect on the line
profile \citep{haiman02,santosm03,loeb04}.  In our model, we expect
this absorption to be less significant, because most galaxies are
embedded in \ion{H}{2} regions with large sizes, even relatively
early in the reionization process.

Finally, in this paper we have considered only the simplest
reionization histories with a single type of source.  Many models for
reconciling the quasar and CMB data on reionization require multiple,
distinct generations of sources that cause ``stalling" or even
``double" reionization \citep{wyithe03,cen03,haiman03,sokasian03b}.  Such
histories will of course change the morphology of reionization and
hence modify the 21 cm signal.  Moreover, alternative models of
reionization in which (for example) voids are ionized first yield
different sets of signatures.  In \citet{furl04b}, we use the formalism
developed here to examine how well 21 cm tomography can distinguish
these histories and models.

\acknowledgments 

This work was supported in part by NSF grants ACI AST 99-00877, AST
00-71019, AST 0098606, and PHY 0116590 and NASA ATP grants NAG5-12140
and NAG5-13292 and by the David and Lucille Packard Foundation
Fellowship for Science and Engineering.


\end{document}